\def\f{\frac}
\def\b{\begin{eqnarray}}
\def\e{\end{eqnarray}}
\def\rLL{1}
\def\rC{2}
\def\rF{3}
\def\rSh{4}
\def\rH{6}
\def\rHN{7}
\def\rOtt{8}
\def\rDx{9}
\def\rDe{10}
\def\rDp{11}

\documentstyle[aps,prl,multicol,epsf]{revtex}
\begin{document}

\title{Non Perturbative Destruction of Localization \\
in the Quantum Kicked Particle Problem} 

\author{Doron Cohen}

\date{
1991\footnote{This Letter has been submitted to PRL. 
Eventually, the comprehensive paper Ref.[10] that contains 
the reported results, as well as other results, has been published 
first. Consequently there was no longer basis for the publication of 
this Letter.}. Notes added 1999.
} 
   
\address{Department of Physics, 
Technion - Israel Institute of Technology, 
Haifa 32000, Israel}

\maketitle


\begin{abstract}

The angle coordinate of the Quantum Kicked Rotator
problem is treated as if it were an extended coordinate.
A new mechanism for destruction of coherence by noise is analyzed
using both heuristic and formal approach. Its effectiveness
constitutes a manifestation of long-range non-trivial
dynamical correlations. Perturbation theory fails
to quantify certain aspects of this effect. In the perturbative 
case, for sufficiently weak noise, the diffusion 
coefficient ${\cal D}$ is just proportional to the noise 
intensity $\nu$. It is predicted that in some generic cases 
one may have a non-perturbative dependence ${\cal D}\propto\nu^{\alpha}$ 
with $0.35{<}\alpha{<}0.38$ for arbitrarily weak noise. 
This work has been found relevant to the recently studied ionization of
H-atoms by a microwave electric field in the presence of noise~[13]. \\ 
{\bf Note added (a):} Borgonovi and Shepelyansky (Physica D 109, 24 (1997)) 
have adopted this idea of {\em non-perturbative} transport, and 
have demonstrated that the same effect manifests itself in the 
tight-binding Anderson model with the {\em same} exponent $\alpha$. \\
{\bf Note added (b):} The recent interest in the work reported 
here comes from the experimental work by the 
{\em Austin} group (Klappauf, Oskay, Steck and Raizen,  
PRL 81, 1203 (1998)), and by the 
{\em Auckland} group (Ammann, Gray, Shvarchuck and Christensen, 
PRL 80, 4111 (1998)).  In these 
experiment the QKP model is realized literally. 
However, the novel effect of {\em non-perturbative transport}, reported 
in this Letter, has not been tested yet. 
\end{abstract}

\begin{multicols}{2}
 
The most striking manifestation of quantum mechanical effects
on classical chaos is dynamical localization which leads to
suppression of chaos. Consider for example a particle that
is confined to move in a one dimensional space whose length 
is $L$ and that is subject to
 a kicking potential with period $T$. Classically, 
the motion of the particle becomes ergodic in space but diffusive
in momentum$^{\rLL}$. Thus, the kinetic energy of the particle grows
like $\langle p^2\rangle \sim D_0t $ , where the diffusion coefficient
$D_0$ depends on the strength of the kicking potential.
Quantum mechanically it is found that diffusion in momentum
is suppressed$^{\rC}$. This is due to localization of the Floquet  
eigenstates in momentum$^{\rF}$. A standard argumentation$^{\rSh}$ leads 
to the following expression for the localization length
\b 
\ell \ = \ \f{2\pi}{L}\hbar\xi \ = \ \f{TL}{2\pi\hbar}D_0
\e
where $\ell$ is measured in units of $p$ while $\xi$ is the
dimensionless localization length. The prototype problem for
the investigation of dynamical localization is the Quantum
Kicked Rotator (QKR) Problem$^{\rC,\rF}$. In this problem the particle
is kicked by a {\bf cos}$x$ potential and periodic boundary conditions
are imposed over $[0,2\pi]$. However, we may consider $x$ to be
an extended variable and impose periodic boundary conditions
over $[0,2\pi{\cal M}]$ where ${\cal M}$ is an integer and the
limit ${\cal M}\rightarrow\infty$ is taken. We obtain then 
a new problem to be entitled 'The Quantum Kicked Particle (QKP)
Problem'. It is {\em not} correct to use (1) with 
$L=2\pi{\cal M}$ to obtain $\ell=T\f{D_0}{\hbar}{\cal M}
\rightarrow\infty$ since due to the translational symmetry of
{\bf cos}$x$ the localization length $\ell$ is the same as in
the QKR problem (${\cal M}=1$) irrespective of ${\cal M}$.  
However, it is evident that any dislocation in the periodic structure
of the kicking potential will result in $\ell\rightarrow\infty$.
Therefore we expect localization in the QKP problem to be extremely
sensitive to any generic perturbation. We shall discuss in this
letter the effect of noise on localization in the QKP problem.  
In conclusion we shall explain why this problem should
be considered a prototype example$^{5}$ for the  
recently studied noise-induced diffusive-ionization of a
highly excited H-atom that is subject
to a monochromatic microwave electric field$^{\rH,\rHN}$. \\

We are considering in this letter the quantized version of the
classical standard map with noise, namely
\b 
x_{t} & = & x_{t-1} + p_{t-1}  \nonumber \\
p_{t} & = & p_{t-1} + K\sin x_{t} + f_{t} 
\e
It is implicit that the dynamical behavior should be averaged over 
realizations of the sequence $f_t$ such that $\langle f_t\rangle=0$ and
\b 
\langle f_{t}f_{t'}\rangle \ = \  \nu\delta_{t,t'}
\e
Following Ott, Antonsen and Hanson$^{\rOtt}$ we assume that the one-step
propagator that generates this map is 
\b 
\hat{U} \ = \ \exp\left[-\f{i}{\hbar}(K\cos\hat{x} + \hat{V}_{int})\right]
\ \exp\left[-\f{i}{\hbar}\f{1}{2}\hat{p}^{2}\right]
\e
where ${V}_{int}$ is the interaction term with the noise source.
Consider first the standard QKR case in which $x$ is an angle
variable. The interaction 
term must respect then the $2\pi$ spatial periodicity of $x$.  
Possible choices that correspond to the classical map (2)
are $\hat{V}_{int}=\sqrt{2\nu}\sin (\hat{x}+\varphi(t))$ where
$\varphi(t)$ is a random phase$^{\rOtt}$, 
and$^{\rDx}$ $\ \ \ \hat{V}_{int}=\int d\varphi
f_{\varphi}(t)\sqrt{2}\sin (\hat{x}+\varphi)$ where $f_{\varphi}(t)$
satisfies 
$\langle f_{\varphi}(t) f_{\varphi'}(t')\rangle = \nu\f{1}{2\pi}
\delta(\varphi{-}\varphi')\delta_{t,t'}$. 
It may be shown that this QKR model is not sensitive to the 
detailed form of the interaction term$^{10}$. 
In the QKP problem the map (2) describes
the time evolution of a {\em particle}. A generic interaction 
term with the external noise source is not expected to respect
the $2\pi$ spatial periodicity of the kicking potential.  
We may assume then e.g. a linear coupling scheme 
$\ \  V_{int}=-f_t\hat{x}$  where $f_t$ satisfies (3). 
We shall see that in this QKP model the dynamical behavior is
significantly different compared with the QKR model though
both models correspond to the same map (2).
From now on it is assumed that $1\ll K$ 
which is  the usual condition for being
in the classically-chaotic regime of the standard map. \\

In the presence of strong noise diffusion in momentum is 
classical-like$^{\rOtt}$ with coefficient $\f{1}{2}K^2+\nu$. If
the noise is weak then classical-like diffusion lasts a characteristic
time $t^\ast\approx2\xi$ and then a crossover to slower diffusive
behavior is observed. The asymptotic diffusion coefficient
is defined as follows:
\b 
{\cal D} \ = \ \lim_{t\rightarrow\infty} \ 
\f{\langle\langle (p(t)-p(0))^2\rangle\rangle }{t} 
\e
where $\ll \ \ \gg$ denotes quantum statistical average over initial
conditions and noise realizations (see Ref.\rDe \ \ for further details). 
 In the absence of noise ${\cal D}=0$ due to the localization 
effect. 
We shall use now a heuristic picture in
order to determine ${\cal D}$ in the presence of weak noise. Next 
we shall introduce a formal treatment and the limitations of
both approaches will be pointed out. \\

A good way to gain insight of the effect of noise on coherence
is to use Wigner's picture of the dynamics$^{\rDp}$. Wigner's function
$\rho (x,p)$ is defined on $[0,2\pi{\cal M}]\times\f{\hbar}{2{\cal M}}
{\cal Z}$ where ${\cal Z}$ are the integer numbers. Assuming that
the particle is prepared in a $\hat{U}$-eigenstate, Wigner's
function has details on spatial scale $\f{1}{\xi}$ indicating
a superposition of $\xi$ momentum eigenstates. The effect of noise
is to smear fine details of Wigner's function and thus to turn the 
superposition into a mixture$^{\rDp}$. The coherence time $t_c$ in the
QKR problem is simply the time it takes for the noise to 'mix'
neighboring momenta$^{\rOtt}$ on momentum scale $\hbar$ , namely
$t_{c}^{QKR} = (\hbar^2\f{1}{\nu})$ , while in the QKP problem a
shorter time scale exists, namely $t_{c}^{QKP} =
 (\f{1}{\xi^2}\f{1}{\nu})^{\f{1}{3}}$ which is the time it takes 
to {\em spread} over spatial scale $\f{1}{\xi}$. This spreading
is absent in case of a rotator since it is associated with the
noise-induced diffusion in the non-discrete momentum space. This
diffusion is $\delta p \propto t^{\f{1}{2}}$  while the associated
spreading is $\delta x \propto t^{\f{3}{2}}$.
It is important
to note that implicit in this heuristic picture is the underlying
assumption that the kicks do not affect significantly the 
coherence time. This assumption has been shown to be incorrect
in case of the QKR problem
if the noise possesses long range correlations$^{\rDp}$. Actually, we shall
see that in the QKP problem the situation is similar, though the
heuristic picture gives the right qualitative behavior. \\

\begin{figure}
\epsfysize=2.2in 
\epsffile{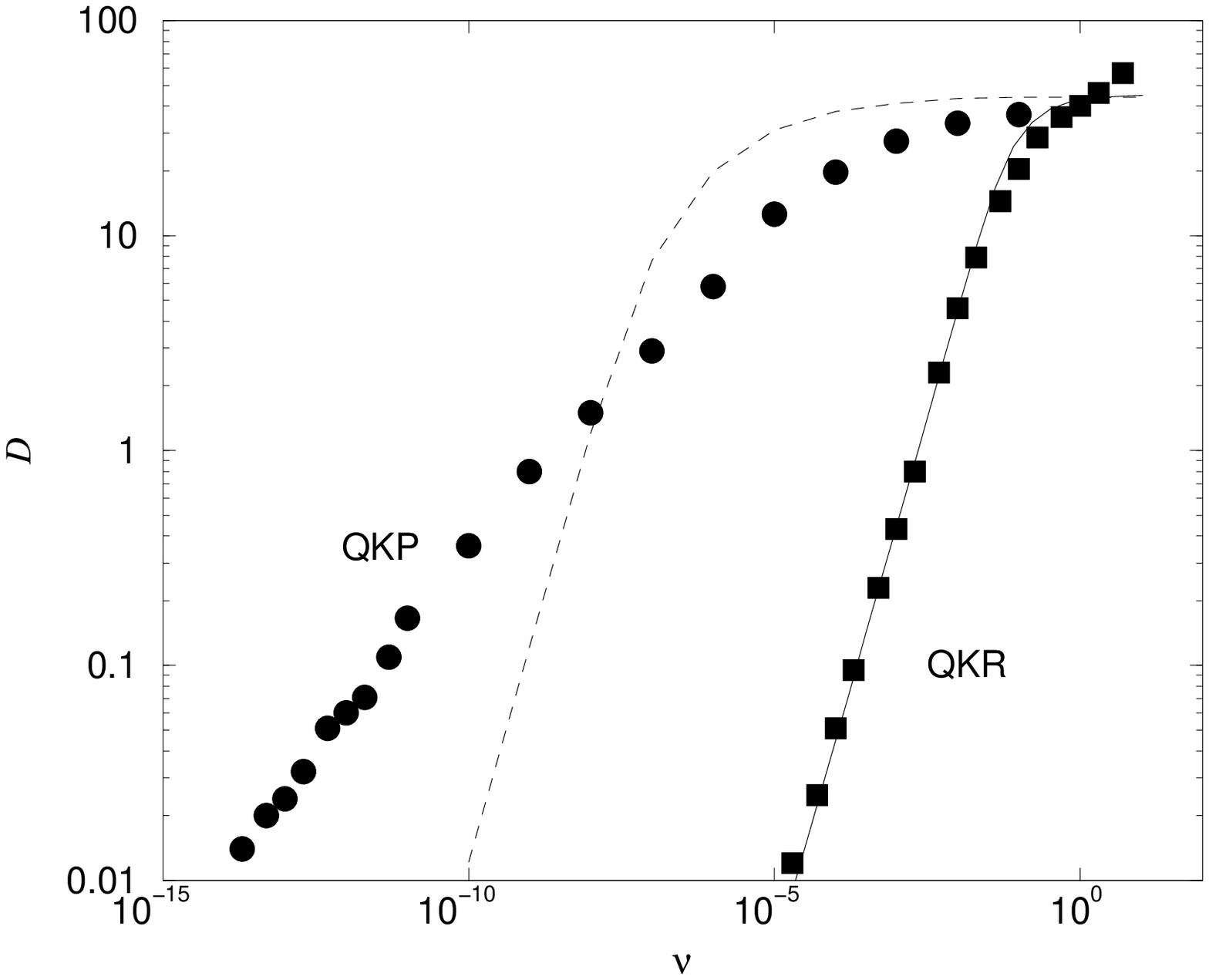}

\noindent
{\footnotesize {\bf FIG. 1.}
\ The diffusion coefficient ${\cal D}$
that has been determined by simulations as a function
of $\nu$ in both the QKR (filled squares) and the QKP (filled circles) 
problems. The parameters are 
$K=10$ and $\hbar=2\pi\times 0.3/(\sqrt{5}-1)$. 
Average over 100 realizations of the noise has been taken 
and the number of iterations was up to $10^4$ for each realization.
The smooth curves has been calculated using (10) and assuming
the short time behavior in (7) to hold throughout the whole 
time domain. Either (8) or (9) were substituted. There are
no fitting parameters. The numerical results in case of 
the QKP model (circles) definitely deviates from the calculated 
perturbative result (dashed line). Rather than having a perturbative 
behavior ${\cal D}\propto\nu$ we have a non-perturbative 
behavior ${\cal D}\propto\nu^{\alpha}$ with $0.35<\alpha<0.38$. 
This non-perturbative behavior persists for arbitrarily weak noise.}
\end{figure}

We proceed now to estimate ${\cal D}$. One may try to use the 
heuristic diffusion picture that is implicit in the work by
Ott Antonsen and Hanson$^{\rOtt}$. It is argued that for weak noise 
($t^{\ast}\ll t_c$) the diffusion process in momentum space is
similar to a random walk on a grid with spacing $\hbar\xi$ 
and hopping-probability $\f{1}{t_c}$. The diffusion coefficient 
in the presence of weak noise is therefore of the order 
$(\hbar\xi)^2\f{1}{t_c}$ which upon using (1) leads to
\b 
{\cal D} \ \approx \ \f{t^{\ast}}{t_c}D_0 \ \ \ .
\e
It follows that ${\cal D}\propto\nu^{\alpha}$ with $\alpha=1$
for the QKR and $\alpha=\f{1}{3}$ for the QKP. In Figure 1 
the results of a numerical experiment are presented$^{13}$.
The observed behavior is
indeed  $\alpha=1$ for the QKR but $0.35<\alpha<0.38$ for the 
QKP  which deviates slightly from
the heuristic value $\alpha=\f{1}{3}$.
We turn now to a formal analysis  of the problem
to overcome the natural limitations of the above
heuristic picture. We shall try to  explain the origin for
the deviation from the heuristic
result in case of the QKP, but we shall 
see that leading order perturbation theory       
cannot be trusted if we want to quantify this deviation. \\
 
In the absence of noise one may define the dispersion (energy)
function $E(t)\equiv\langle\langle (p(t)-p(0))^2\rangle\rangle$. 
This function is related
to the momentum autocorrelation function$^{10}$ via 
$E(t)=2(C_p(0)-C_p(t))$. Its time derivative will be denoted by 
$D(t)$, and has the asymptotic value ${\cal D}=0$ due to the
localization effect. It has been found$^{10}$ that dynamical correlations 
in the QKR problem decay exponentially on time scale $t^*$ while
on longer time scale a slower power-law decay is observed.
Consequently
\b
D(t) \ = \ \left\{ \begin{array}{cl}
D_0 \ e^{-\f{t}{t^*}} & \ \ \ \mbox{for} \ \ t<{\cal O}(t^*) \\
cD_0 (\f{t^*}{t})^{1+\beta} & \ \ \ \mbox{for} \ \ {\cal O}(t^*)<t
\end{array} \right.   
\e
with $D_0\approx\f{1}{2}K^2$, \ $t^*\approx2\xi$, \ 
$\beta\approx0.75$ and $c\approx0.5$. More details including
analytical considerations may be found in Ref.10. \\

In the presence of noise coherence is destroyed.
The decay probability $P(t)$ of a quasienergy eigenstate as a
function of time may be calculated using leading order perturbative 
calculation$^{\rDx,\rDe}$.
For the QKR the decay rate is constant,namely
\b
\dot{P}(t) \ = \ \f{1}{\hbar^2} \nu \ \ \ .
\e
For the QKP one obtains
\b
\dot{P}(t) \ = \ \f{1}{\hbar^2} \nu \ 
\sum_{\tau=-t}^{t} C_p(\tau) (t-|\tau|) \ \ \ .
\e
For $t^*\ll t$ the behaviour is roughly 
$P(t)\approx\nu\xi^{2+\beta}t^{3-\beta}$ in the latter case.
These results hold as long as $P(t)\ll1$. However,
if we assume that $P(t)$ is a function of a single scaled variable
$\f{t}{t_c}$ then the perturbative result suggests that
for the QKR problem
$t_{c}^{QKR} = (\hbar^2\f{1}{\nu})$
which is the inverse of the decay rate and agrees with our
heuristic expectation. For the QKP one obtains
$t_{c}^{QKP}\approx(\f{1}{\xi^{2+\beta}}\f{1}{\nu})^{\f{1}{3-\beta}}$
that coincide with the heuristic result only if we assume very strong
dynamical correlations ($\beta\rightarrow 0$) which is not correct
since $\beta\approx 0.75$.   
Note that the latter results are as exact as leading order perturbation
theory permits. \\

We consider now diffusion in presence of weak noise
($t^*\ll t_c$) using a formal approach. A derivation$^{10}$ which
is based on leading order perturbation theory leads to the 
result
\b
{\cal D} \ \approx \ \sum_{t=0}^{\infty} \dot{P}(t)D(t) \ \ \ .
\e
This expression may be trusted only if it is dominated by the
short-time terms (those with $t\ll t_c$). This would be always
the case if dynamical correlations possesed a short-range nature.
Specifically, if $D(t)$ decayed exponentially on the relatively
short time scale $t^*$, then the sum in (10) would be
dominated by the terms in its head whose number is of the
order $t^*$. Evidently ${\cal D}$ should be proportional then
to the intensity of the noise. A non-trivial dependence of
the form ${\cal D}\propto\nu^{\alpha}$ with $\alpha\ne1$ is
therefore a manifestation of long range dynamical correlations.
The sum in (10) is necessarily dominated then by the
long-time terms and consequently the perturbative estimate
for ${\cal D}$ cannot be trusted any more. \\

In case of the QKR model one easily finds that in spite of the
power law decay of the long-time terms, yet the sum (10) is
dominated by the short time terms. Substitution of (7) and
(8) into (10) leads then to the heuristic formula (6).
Figure 1 illustrates comparision of the numerical results
(filled squares) with the analytical estimate (smooth curve,  
no fitting parameters). We turn now to discuss the QKP
case. Here the behaviour of the terms in the sum (10) that
satisfy $t^*\ll t\ll t_c$ is
\b
\dot{P}(t)D(t) \ = \ 
(3{-}\beta)cD_0\f{(t^*)^{1+\beta}}{(t_c)^{3-\beta}} \
t^{1-2\beta} \ \ \ ,
\e
where we have used (7) and (9). This behaviour 
(provided $\beta\le1$) indicates that most of the contribution 
to ${\cal D}$ in (10) comes from the long-time terms with 
$t$ which is of the order $t_c$. This observation is supported
by the comparision of the numerical$^{12}$ results (Figure 1, filled
circles) with analytical estimate that takes into account
only the short-time contribution (dashed curve, no fitting parameters). 
One may try to use the following extrapolation scheme in order to
estimate the dominant contribution of the long-time
terms to the diffusion coefficient : \ \ (a) To assume that
nevertheless (10) holds, \ \ (b) To assume that (9)
holds for $t\le t_c$ while $\dot{P}(t)=0$ for $t_c<t$. 
One obtains then 
${\cal D}=c'(\f{t^*}{t_c})^{1+\beta}D_0$
instead of (6) where $c'$ is a prefactor of order unity. 
Consequently ${\cal D}\propto\nu^{\alpha}$ 
with $\alpha=\f{1+\beta}{3-\beta}$.
This result differs for $0<\beta$ from the heuristic one. It
predicts that $\alpha$ is larger then $\f{1}{3}$,   
namely $\alpha\approx 0.78$.  Unfortunately, the latter
value does not agree with the numerical results which leads
to the conclusion that leading order perturbation theory is
not sufficient in order to obtain  a quantitive description
of the effect. \\

The QKP problem is a prototype example that illustrates destruction
of coherence via a {\em spreading mechanism}. This mechanism is
opperative in systems where the noise does not respect
a {\em symetry} that is responsible for the localization.
In the QKP problem, due to translational symetry of the kicking
potential, only states that have finite momentum separation are
coupled by the kicks. This
feature is shared by the recently studied highly excited
H-atom that is subject to a monochromatic microwave electric
field$^{\rH}$. The high energy levels of the
undriven H-atom are very  dense, but 
{\em only photon-distant states are coupled by the interaction
with the field}. It follows that this problem
reduces locally to a 
generelized QKP-problem with finite ${\cal M}$
rather then QKR-problem. Generic
noise will induce diffusion to neighbouring levels 
that play no significant role in the dynamics if the 
noise is absent. If the time scale  that is required 
for this diffusion is much less than $t_c^{QKP}$   
then we may expect that the 
spreading mechanism for destruction of coherence will become effective. 
Our results therefore sugests that if the H-atom is prepared
in a {\em very} high excited state, then a new behaviour which
is different from the one that has been reported
in Ref.\rHN \ \  may be found. Namely, the ionization time will
not be in general inverse proportional to the variance of the noise.
This subject obviously deserves a 
systematic study. Indeed Fishman and Shepelyansky$^{13}$ 
have considered the effect of noise on ionization and pointed
out that there are indications for the manifestation of the 
non-perturbative mechanism in experiments on Rubidium atoms
that have been carried out recently by the Munich group$^{14}$. \\ 

{\bf Note added (c):} Further discussion of the 
perturbative and the non-perturbative mechanisms for 
destruction of cohernce, in a more general context, 
may be found in the paper 
"Quantal Brownian Motion - Dephasing and Dissipation" 
(D. Cohen,{\it J. Phys. A {\bf 31}}, 8199 (1998)).
It should be emphasized that the essential ingredient 
for the manifestation of the non-perturbative mechanism  
is the possibility for exchange of relatively small quanta 
of momentum  between the particle and the environment. 
It should be possible to realize such type of noise in 
eg Raizen's experiments by introducing a noisy 
field with small $q$ components 
($q=$ wavenumber in the relevant direction).  
The emphasis on `symmetry breaking' in the above 1991 version 
of the Letter is somewhat misleading. \\


I thank S. Fishman for useful criticism , D.L. Shepeliansky
for fruitful discutions and T. Dittrich, F.M. Izrailev and 
E. Shimshony for interesting conversations.
I also thank R. Graham and F. Haake for thier hospitality
in Universitat-GHS-Essen.    This
work was supported in part by the U.S.-Israel Binational Science
Foundation (BSF), and by the European Science Foundation (ESF). \\

\rule{8cm}{.01in} \\
 
\begin{enumerate}

\item A.J. Lichtenberg and M.A. Lieberman, {\it Regular and Stochastic
Motion}, (Springer, Berlin, 1983).
 
\item G. Casati, B.V. Chirikov, F.M. Izrailev and J. Ford, in
{\it Stochastic Behaviour in classical and Quantum Hamiltonian Systems},
Vol. 93 of {Lecture Notes in Physics}, edited by G. Casati and
J. Ford (Springer, N.Y. 1979), p. 334.
 
\item S. Fishman, D.R. Grempel and R.E. Prange, Phys. Rev. Lett. {\bf
49}, 509 (1982). D.R. Grempel, R.E. Prange and S. Fishman, Phys. Rev.
A {\bf 29}, 1639 (1984). S. Fishman, R.E. Prange, M. Griniasty, Phys.
Rev. A {\bf 39}, 1628 (1989). S. Fishman, D.R. Grempel and R.E.
Prange, Phys. Rev. A {\bf 36}, 289 (1987).

\item B.V. Chirikov, F.M. Izrailev and D.L. Shepelyansky,
Sov. Sci. Rev. {\bf 2C}, 209 (1981).      
D.L. Shepelyansky, Physica {\bf 28D}, 103 (1987).

\item I thank D.L. Shepelyansky for suggesting this connection.
 
\item J.E. Bayfield and P.M. Koch, Phys. Rev. Lett. {\bf 48},
711 (1982). J.G. Leopold and I.C. Percival, J. Phys. B {\bf 12},
709 (1979). R. Blumel and U. Smilansky, Z. Phys. D {\bf 6},
83 (1987). G. Casati, I. Guarneri and D.L. Shepeliansky, IEEE J.
Quant. Elect. {\bf 24}, 1240 (1988).  

\item R. Blumel, R. Graham, 
L. Sirko, U. Smilansky, H. Walther and K. Yamada, 
Phys. Rev. Lett. {\bf 62}, 341 (1989).
 
\item E. Ott, T.M. Antonsen Jr. and J.D. Hanson, Phys. Rev. Lett.
{\bf 53}, 2187 (1984).
 
\item D. Cohen, preprint (1991). \\ 
{\bf Note:} Published in J. Phys. A {\bf 27}, 4805 (1994).

\item D. Cohen, Phys. Rev. A {\bf 44}, 2292 (1991).                 

\item D. Cohen, phys. Rev. Lett. {\bf 67}, 1945 (1991) ; 
Phys. Rev. A {\bf 43}, 639 (1991).
 
\item I thank D.L. Shepelyansky for useful idea concerning 
these simulations.

\item S. Fishman and D.L. Shepelyansky, Europhys. Lett.,
{\bf 16}(7), pp. 643-648 (1991).

\item A. Buchleitner, R. Mantegna and H. Walther, presented
at the Marseille Conference on Semiclassical Methods in 
Quantum Chaos and Solid State Physics, and to be published.

\end{enumerate}
\end{multicols}
\end{document}